\documentclass[12pt]{article}

\newenvironment{wileykeywords}{\textsf{Keywords:}\hspace{\stretch{1}}}{\hspace{\stretch{1}}\rule{1ex}{1ex}}

\usepackage{xr}
\makeatletter
\newcommand*{\addFileDependency}[1]{% argument=file name and extension
  \typeout{(#1)}
  \@addtofilelist{#1}
  \IfFileExists{#1}{}{\typeout{No file #1.}}
}
\makeatother

\newcommand*{\myexternaldocument}[1]{%
    \externaldocument{#1}%
    \addFileDependency{#1.tex}%
    \addFileDependency{#1.aux}%
}

\myexternaldocument{SI}

\usepackage{amsmath,amssymb}
\usepackage{graphicx}% Include figure files
\usepackage{color}% Include colors for document elements
\usepackage{dcolumn}% Align table columns on decimal point
\usepackage{bm}% bold math
\usepackage[numbers,super,comma,sort&compress]{natbib}
\usepackage{multirow}
\usepackage{xcolor}
\usepackage[normalem]{ulem}

\definecolor{background-color}{RGB}{255,255,255}

\title{Ab Initio Investigation of Excited States and Absorption Spectra of Small Mercury Clusters (Hg$_{2-5}$)}
\author{Michal Novotn\'{y}\thanks{Department of Physics, Faculty of Science, University of Ostrava, 30.~dubna 22, 701 03 Ostrava, Czech Republic, michal.novotny@osu.cz, frantisek.karlicky@osu.cz}, Franti\v{s}ek Karlick\'{y}\footnotemark[1]}

\begin{document}

\maketitle

\begin{abstract}
We report a systematic \textit{ab initio} investigation of the excited states and absorption spectra of small mercury clusters (Hg$_{\mathrm{2-5}}$). Benchmarking multi-reference configuration interaction (MR-CISD) and equation-of-motion coupled-cluster (EOM-CCSD) methods shows that EOM-CCSD/aug-cc-PVTZ-PP reproduces atomic excitation energies within $\pm$500 cm$^{-1}$ of experiment, whereas MR-CISD deviates by about $\pm$1100 cm$^{-1}$. For the mercury dimer, EOM-CCSD potential energy curves yield absorption bands in excellent agreement with experimental transitions, while asymptotic excitation energies closely match atomic references. Extending this methodology, we present the first systematic \textit{ab initio} study of absorption spectra of Hg$_{\mathrm{3}}$, Hg$_{\mathrm{4}}$, Hg$_{\mathrm{5}}$, which display broad bands in the ranges of 37000-60000 cm$^{-1}$. The consistent agreement with experimental benchmarks and diagnostics confirming single-reference character validates EOM-CCSD as a robust approach for describing excited states and optical spectra of mercury clusters.
\end{abstract}

\begin{wileykeywords} 
Hg, mercury, cluster, dimer, van der Waals cluster,  EOM-CCSD, EOM-CC, EOMCCSD, EOMCC, CCSD(T), ab-initio, spectra, absorption spectra
\end{wileykeywords}

\clearpage

%*****************Graphical Table of Contents******************** THIS IS MANDATORY *******************

%\begin{figure}[ht]
%\centering
%\colorbox{background-color}{
%\fbox{
%\begin{minipage}{1.0\textwidth}
%\includegraphics[width=50mm,height=50mm]{other/Graphical_abstract.eps} 

%This work establishes a reliable \textit{ab initio} framework for excited states of small mercury clusters (Hg$_{2}$-Hg$_{5}$). Benchmarking shows EOM-CCSD accurately reproduces atomic and dimer excitations and yields absorption spectra in quantitative agreement with experiment. Extending to larger clusters reveals broad high-energy bands and consistent electronic structure trends. The results demonstrate that single-reference methods and effective core potentials can robustly describe optical properties in these relativistic systems, enabling predictive modeling of mercury cluster spectroscopy.
%\end{minipage}
%}
%}
%\end{figure}

% makes references listed with 1., 2., etc.  
  \makeatletter
  \renewcommand\@biblabel[1]{#1.}
  \makeatother

\renewcommand{\baselinestretch}{1.5}
\normalsize

\clearpage

%%%MAIN TEXT%%%%
%===================================
\section{Introduction}
\label{intro}
%===================================
Mercury clusters represent a unique class of systems that bridge the gap between molecular and bulk metallic behavior and provide a platform for studying the evolution of bonding and electronic properties in heavy, relativistic elements. Among these, small neutral mercury clusters (Hg$_2$–Hg$_5$) serve as prototypical systems for understanding the evolution from weakly bound van der Waals complexes to metallic behavior~\citep{Brechignac1985, Brchignac1988, Pastor1988}. In particular, stabilization of the 6s orbital and its weak hybridization with the 6p orbitals in small clusters hinder the development of delocalized metallic bonding~\cite{Kitamura2006, Garcia1991}. As cluster size increases, this behavior changes gradually, with electronic structure transforming from a van der Waals through covalent-like regimes to metallic bonding only at much larger sizes ($N > 30$–$50$)~\cite{Rademann1987, Harbour1971, BCabaud1980}. Yet for the smallest clusters ($N = 2$–$5$), these effects are especially pronounced and amenable to high-level theoretical treatment.

Experimentally, the optical properties of small mercury clusters remain largely unexplored, particularly for neutral species in the low-$N$ regime. While extensive work has focused on plasmonic responses in larger metal clusters of alkali and coinage metals~\cite{Yannouleas1990, Fallgren1990}, divalent metals such as mercury have received relatively less attention. Nevertheless, the sharp size-dependent evolution of optical absorption in divalent metal clusters has begun to emerge as an important probe of their changing electronic structure~\citep{Rademann1987, Rademann1992}. For mercury, absorption studies not only complement insights from ionization potentials and autoionization spectra, but also provide access to complex excited-state behaviour associated with  6s$^2$–6s6p transitions, which are sensitive to both structure and relativistic interactions~\citep{Kitamura2006}.

Theoretical high-level \textit{ab initio} calculations on mercury clusters are notoriously challenging due to the need for proper treatment of relativity, electron correlation, and excited-state mixing~\citep{Kitamura2006, Kitamura20062}. For these reasons, there are only a few studies concerned with \textit{ab initio} calculations of neutral mercury clusters~\cite{Vitek2022, Sarmanova2019}, more specifically of the mercury dimer. Previous theoretical works of note concerned with the neutral mercury dimer are Czuchaj et al.~\cite{Czuchaj1997}, Kullie~\cite{Kullie2014}, and Krosnicki et al.~\cite{Kronicki2015}, and Balasubramanian et al.~\cite{Balasubramanian1992}, which we will be contrasting with our results in later sections. For the excited states of larger clusters, no \textit{ab initio} studies exist, but semi-empirical approaches based on diatomics-in-molecules models have been employed with some success\cite{Calvo2012, Cosi2018}. 

This paper aims to assess if a multi-reference treatment of small mercury clusters is necessary for the calculation of their absorption spectra, as well as determine the level of theory required to properly account for relativistic effects. The motivation is to provide highly accurate and computationally accessible potential energy curves (PEC) for the lower and higher excited states of the mercury dimer and establish an accurate and efficient methodology to study larger systems. In this paper, we also present \textit{ab initio} level excitation energies and absorption spectra of Hg$_{\mathrm{1-5}}$, which for the larger clusters (3-5) have not been reported in any previous works.
%===================================
\section{Methods}
\label{methods}
%===================================
\subsection{Basis sets}
In the calculations, we used correlation-consistent basis sets: aug-cc-PVXZ-PP (X = D, T, Q, 5) with effective core potentials (ECP) as well as the corresponding all-electron aug-cc-PVXZ-DKH basis sets and the
%Atomic Natural Orbitals Relativistic Core–Correlated 
ANO-RCC-XZ (X = D, T, Q)~\cite{Pritchard2019} basis sets. 
\subsection{Ground state calculations}
Calculations of ground states were performed using coupled-cluster with singles, doubles, and perturbative triples (CCSD and CCSD(T)), multi-reference configuration interaction with singles and doubles (MR-CISD), with energies extrapolated using the three-point formula for the complete basis set limit (CBS).

\subsection{Excited state calculations}
We chose equation-of-motion coupled-cluster with singles and doubles (EOM-CCSD) and MR-CISD as the primary methods in this study. MR-CISD was selected as a multi-reference counterpart to EOM-CCSD due to the similar formulation of the coupled-cluster and configuration interaction approaches. In addition, MR-CISD provides explicit excited-state wavefunctions, enabling a consistent evaluation of transition dipole moments and oscillator strengths needed for the calculation of absorption spectra. In contrast, methods such as CAS-NEVPT2 primarily improve excitation energies while transition properties are typically evaluated at the CASSCF level. Consequently, differences between EOM-CCSD and MR-CISD are expected to predominantly reflect nondynamical correlation effects.

\subsubsection{MR-CI}
Calculations employing multi-reference configuration interaction (MR-CI) and state-average complete active space self-consistent field (SA-CAS-SCF) approaches were performed using the ORCA package \cite{RN204, RN155, RN205, RN229, RN84, RN139, RN178, RN21}.

The protocol for calculating MR-CISD excited states was as follows: MP2 natural orbitals (NO) were used as a starting point for an SA-CAS-SCF calculation with a subsequent MR-CISD calculation using the SA-CAS-SCF states as a reference. The MR-CISD routines include the multi-reference Davidson correction for unlinked quadruples to improve size-consistency (MR-CISD+Q). The convergence tolerance on the total energy, orbital gradients, and residuals in the MP2, CAS-SCF, and MR-CISD steps of the calculation was set to 10$^{-12}$ E$_{\mathrm{h}}$.

The active space was constructed by including virtual orbitals with NO occupations greater than 0.02 while maintaining sufficient configurational flexibility within each symmetry. As will be shown in the results, we tested several active space sizes to ensure that non-dynamical correlations were fully accounted for. The tested CAS sizes were (2,13), (2,22) for the mercury atom and (4,8),(4,10),(4,16) for Hg$_{\mathrm{2}}$. We tested several different frozen-core approximation levels with 78, 68, and 46 electrons set as frozen for the mercury atoms. The calculations were performed with a reduced symmetry of D$_{2h}$ since the full rotational symmetry and D$_{\infty h}$ symmetry are not computationally available.

\subsubsection{EOM-CCSD}
The equation-of-motion coupled-cluster (EOM-CC) calculations were performed using the Gaussian 16 Rev. C.01 \cite{g16}. The number of frozen core electrons was set to the default value of 68 electrons under the G2 convention. Due to EOM-CCSD solving a non-Hermitian eigenvalue problem, the number of calculated roots must be increased to ensure accuracy and correct state ordering. The triplet and singlet states were calculated separately with 25 roots included in each calculation. All other parameters were set to use the Gaussian default values.

%\MN{}{add description of fundamental differences between EOM and MRCI: petrurbation/variational, influenced by more root or not, EOM - more robust, evidently doubles in intial guess are not needed}

\subsection{Treatment of relativistic effects}
The Hamiltonian used in both ORCA and Gaussian with ANO-RCC-XZ and aug-cc-PVXZ-DKH basis sets was a one-component spin-free second-order Douglas–Kroll–Hess Hamiltonian. SOC was calculated only for the all-electron basis sets (ANO-RCC-XZ and aug-cc-PVTZ-DKH) using the MR-CISD method. In the SOC calculation, the Breit–Pauli mean-field/effective-potential spin-orbit operator was used together with the Gaussian finite nucleus model of Dyall and Visscher with second-order picture-change effects~\cite{Visscher1997}.
%===================================
\section{Results and Discussion}
\label{results}
%===================================
This paper aims to assess the need for a multi-reference treatment of the excited states of mercury clusters and provide accurate absorption spectra. 
Our methodological analysis will therefore focus on three main aspects: comparison of methods (MR-CISD/EOM-CC), performance of different basis sets (aug-cc-PVXZ/ANO-RCC), and the level of treatment of relativistic effects (ECP/DKH/SOC).
We compared the performance of MR-CISD and EOM-CCSD for each basis set type separately, after which we will discuss the optimal method/basis in general. We will contrast our results with experimental data available for the mercury atom and dimer, as well as contrasting them with previous theoretical studies, mainly from Czuchaj et al.~\cite{Czuchaj1997} and Krośnicki et al.~\cite{Kronicki2015}. In the final part, we will present absorption spectra calculated using the optimized methodology for the larger Hg$_{\mathrm{3}}$, Hg$_{\mathrm{4}}$, and Hg$_{\mathrm{5}}$ clusters.

\subsection{Atom}
While calculating the excitation energies for the mercury atom has been extensively studied, due to it being well experimentally described and calculated to very high precision by previous works~\cite{Srivastava1992, Gleichmann1994,Mosyagin2000}, it serves as a good sanity check for methodologies applied to larger mercury clusters (Hg$_{\mathrm{2-5}}$). Because experimental excitation energies are scarce for Hg$_2$ and unavailable for larger clusters, the Hg atom provides an important benchmark for assessing the methods used here.

In case of the MR-CISD calculations, ORCA uses MP2 orbitals as the starting point. The observations are somewhat similar, although the Dunning basis sets (both all-electron DKH and ECP) produce a virtual space of: (6p 7s 7p 6d 8s 7d 8p) lacking 9s and 9p compared to the Gaussian SCF results. 

\begin{figure}[ht]
    \centering
    \includegraphics[width=0.6\linewidth]{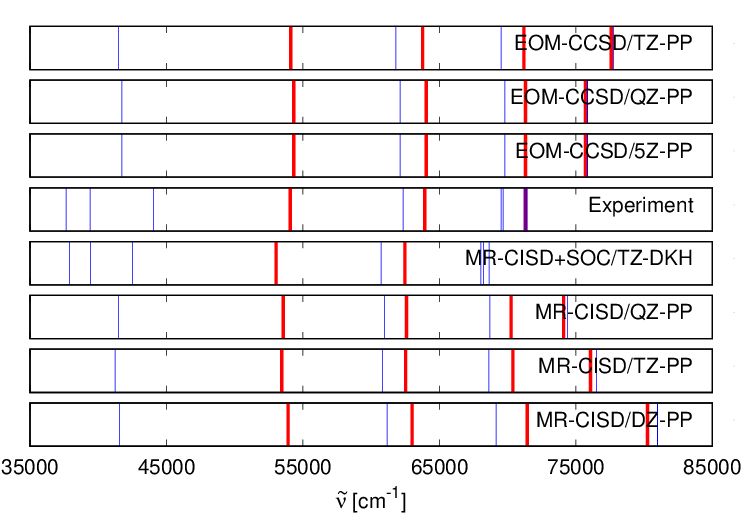}
    \caption{Energy levels of the mercury atom calculated with EOM-CCSD and MR-CISD(2,13) using aug-cc-PVXZ basis sets. Triplet states are coloured blue and singlets in red
    %, the blue dashed line represents the state-averaged first triplet state
    . 
    MR-CISD(2,13)+SOC excitation energies were obtained using the all electron aug-cc-PVTZ-DKH basis set, all other methods employed ECPs.}
    \label{fig:Hg_duning}
\end{figure}

Although not central to the primary objective of this paper, which is to describe the absorption spectrum of Hg$_2$, we will be analysing triplet states as a measure of the robustness of the employed methods while at the same time providing accurate PEC for future spectroscopic investigations. A detailed discussion of the orbital space of the mercury atom is located in the Supplementary Material.

First, we will be discussing the atomic excited states calculated by both methods in the aug-cc-PVXZ-PP basis for which the results are depicted in Fig.~1, and Tab.~S2. Due to most of these calculations being performed without SOC, the $^3$P states became one state-averaged triple-degenerate state, rather than having three separate J-levels.

We first determined the active-space and frozen-core sizes by testing several options. We tested several sizes of the active space (2,8),(2,13),(2,22),(2,44) depicted in Fig.~S1. The (2,13) active space, comprising the occupied 6s and virtual 7s, 6p, 7p, and 6d orbitals, was sufficient to describe all states considered. While the accuracy of the lower lying excited states does not suffer from a smaller active space, (2,8) is still insufficient since it does not include 6d orbitals; it cannot describe the high lying D states. With an increasing CAS size, a slight blue shift in the excitation energies is observed, leading to improved agreement with experimental values. This behavior is expected, as a larger active space provides a more balanced reference wavefunction and can indirectly influence the recovery of dynamical correlation through an expanded set of configurations available for single and double excitations.

We tested a frozen core of 78 electrons as used in Czuchaj et al.~\cite{Czuchaj1997} down to 60 electrons as depicted in Fig.~S2. As expected, a frozen core of 78 electrons is insufficient since it does not account for the d-electron correlation. Accounting for this correlation results in an overall blue shifts the whole spectrum closer to the experimental reference as well as producing a correct order and energy of the excited states. While nonsensical from the point of view of electronic structure, freezing an increasing number of 5d electron pairs shows us the significance of their contribution to the correlation energy, as well as the negligible contribution of the 5p and 5s electrons to the total correlation energy.

After setting the frozen core size to 68 electrons and choosing a (2,13) active space, we proceeded to calculate the excited states of the mercury atom shown in Fig.~\ref{fig:Hg_duning} and table Tab.~S2. The data show that, in general, the lower-lying excited states (first two singlets and triplets) are well described by both methods and most basis sets, as we assumed from our orbital analysis. The only incorrect energy spectrum was produced by EOM-CCSD using aug-cc-PVDZ, which has all triplet states blue shifted by more than 15000 cm$^{-1}$, which was therefore excluded from our analysis.

From orbital analysis, we have seen that many excited states, S and P states in particular, include excitations into relatively high-lying orbitals (8s, 9s, 8p, etc.) with non-negligible coefficients, reinforcing the need for a large basis set for accurate description of even lower lying excitations. We expect this to be the main reason for the blue shift compared to the experiment of the three highest examined excited states.

Both methods achieve good agreement with experimental data with EOM-CCSD (5Z) deviating on average by $\pm$ 500 cm$^{-1}$ for non-SOC states ($^{1}\mathrm{P}_{1},^{1}\mathrm{S}_{1},^{1}\mathrm{P}_{2},^{1}\mathrm{D}_{1}$) and MR-CISD faring noticeably worse $\pm$ 1100 cm$^{-1}$. A probable source is how well these methods describe dynamical correlation. MR-CISD is employed in systems where the contribution of non-dynamical correlation is comparable to dynamical correlation. EOM-CCSD with its CC ansatz and infinite order energy contribution manages to account for more dynamical correlation energy, which is the main component in the atomic spectrum. How much and which electrons contribute to this is illustrated in the Supporting Information with the evolution of the spectrum with increasing number of frozen electrons. Previous studies, like Balasubramanian et al.~\cite{Balasubramanian1992} and   Czuchaj et al., had to employ a strongly constrained approach (frozen core of 78 electrons) due to the computational limitations of the time. Surprisingly, they were able to obtain results very close to ours.  Generally, we managed to obtain a better agreement with experiment compared to the previous theoretical work of Balasubramanian et al. (CAS-SCF)~\cite{Balasubramanian1992} and  Czuchaj et al.~\cite{Czuchaj1997} (MR-CISD), Kullie et al.~\cite{Kullie2014} (TD-DFT), and Krossnicki et al.~\cite{Kronicki2015} (CAS-PT2), which should serve as a method validation for use on Hg$_{\mathrm{2}}$. A comparison between this work and other literary sources is shown in Tab.~S1.

While the ECPs provided good agreement with the experiment, we investigated further if an approximate treatment of the scalar and spin-orbit relativistic effect could improve the energy spectrum.

While the ANO-RCC-XZ basis sets (uncontracted or loosely contracted)  are generally recommended for calculations of heavy elements and are specifically designed for multi-reference approaches, our testing (Tab.~S3 and  Fig.~S3) shows significant deficiencies in describing higher-lying excited states. We have therefore performed all subsequent calculations using only the Dunning basis sets.

From the Dunning basis set family, we used the aug-cc-PVTZ-DKH. This basis set provides the most complete result from all the tested methodologies, with split P states, as well as describing all the higher excited states as well. For the highest J of the $^3\mathrm{P}_{1}$, it still falls short of 2000 cm$^{-1}$ compared to experiment, and the splitting of the $^3\mathrm{P}_{2}$ is also insufficient. This is due to the approximate treatment of relativistic effects. For a more accurate description of SOC, a proper two-component method has to be employed. The SOC-calculations serve a qualitative purpose, for checking for state mixing of singlet excited states. SOC is treated by an approximate approach and is by no means appropriate for a detailed discussion of the behavior of triplet and mixed states.

In general, the ECPs manage to be more than sufficient for the description of scalar relativistic effects, and the differences in energy levels arise predominantly from the employed basis set and method.

\subsection{Dimer}
The main objective of this study is to provide both the potential energy curve (PEC) of the mercury dimer and a theoretical absorption spectrum which, to our knowledge, has not yet been calculated using high-level \textit{ab initio} methods. We chose the aug-cc-PVTZ-PP basis set with a frozen core of 68 electrons as an acceptable compromise between accuracy and feasibility for the calculation of the dimer excited states. Hg clusters are characterized by weak bonding and a small HOMO–LUMO gap, which can lead to near-degeneracy of valence configurations. In such cases, a single-reference description may become insufficient, and nondynamical correlation effects can play a non-negligible role. It is therefore necessary to assess whether a multi-reference treatment is required. The limiting factor for these calculations is the size of the active space, where we tested active space sizes of (4,8) and (4,16).

\subsubsection{Orbitals}
\begin{figure}[ht]
    \centering
    \includegraphics[width=0.6\linewidth]{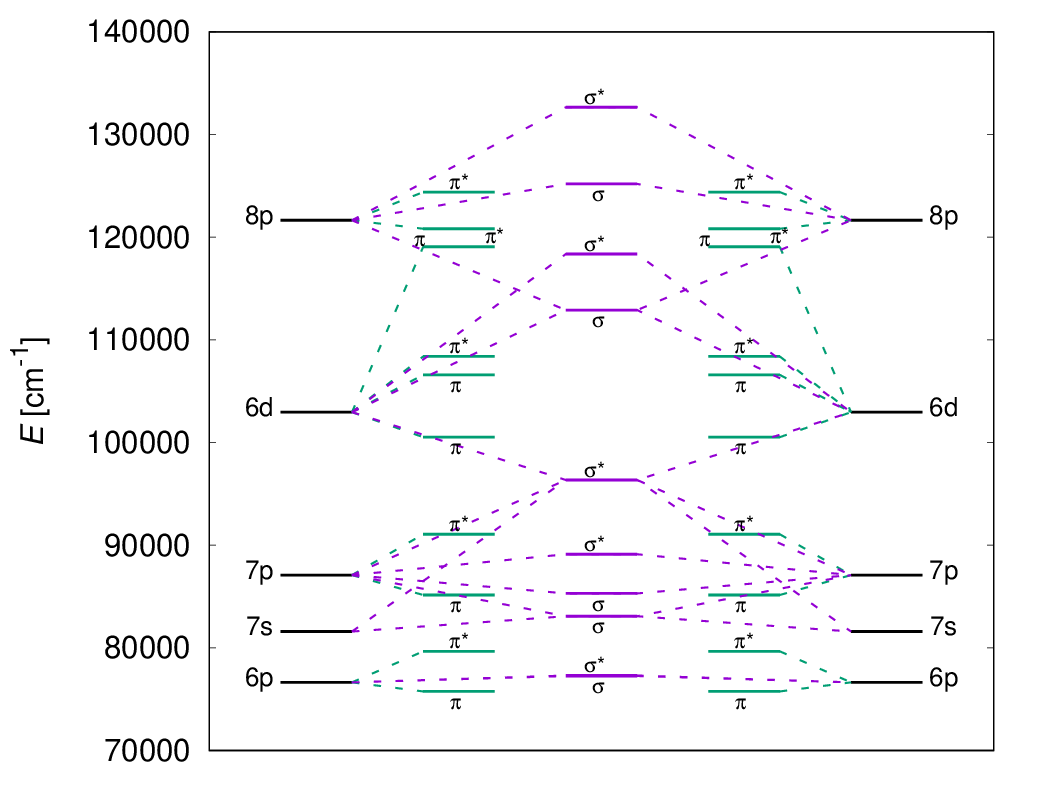}
    \caption{Virtual orbital energies of the mercury dimer and atom in the ground state. For the atom, the Fermi level was set to be equal to the energy of the HOMO orbital, in Hg$_{\mathrm{2}}$, this was aligned with the half point between the HOMO and HOMO$-1$ orbital.}
    \label{fig:Hg2_orbitals}
\end{figure}

The ordering of the molecular orbitals obtained from an MP2 calculation is depicted in Fig.~\ref{fig:Hg2_orbitals}. The excitations considered here originate predominantly from the HOMO and HOMO$-1$, which are mainly composed of the 6s orbitals with a small 5d contribution. The highest MO composed of the 4d-orbitals is 53 700 cm$^{-1}$ (6.67 eV) lower than HOMO$-1$. Combined with the HOMO-LUMO gap of 70 000 cm$^{-1}$ (8.65 eV), these orbitals are not expected to contribute significantly to the UV/vis energy excitations. On the other hand, the HOMO and HOMO$-1$ gap is approximately 12 000  cm$^{-1}$ and is therefore essential to include HOMO$-1$ in the calculations.

\subsubsection{Ground state}
\begin{figure}[ht]
    \centering
    \includegraphics[width=0.6\linewidth]{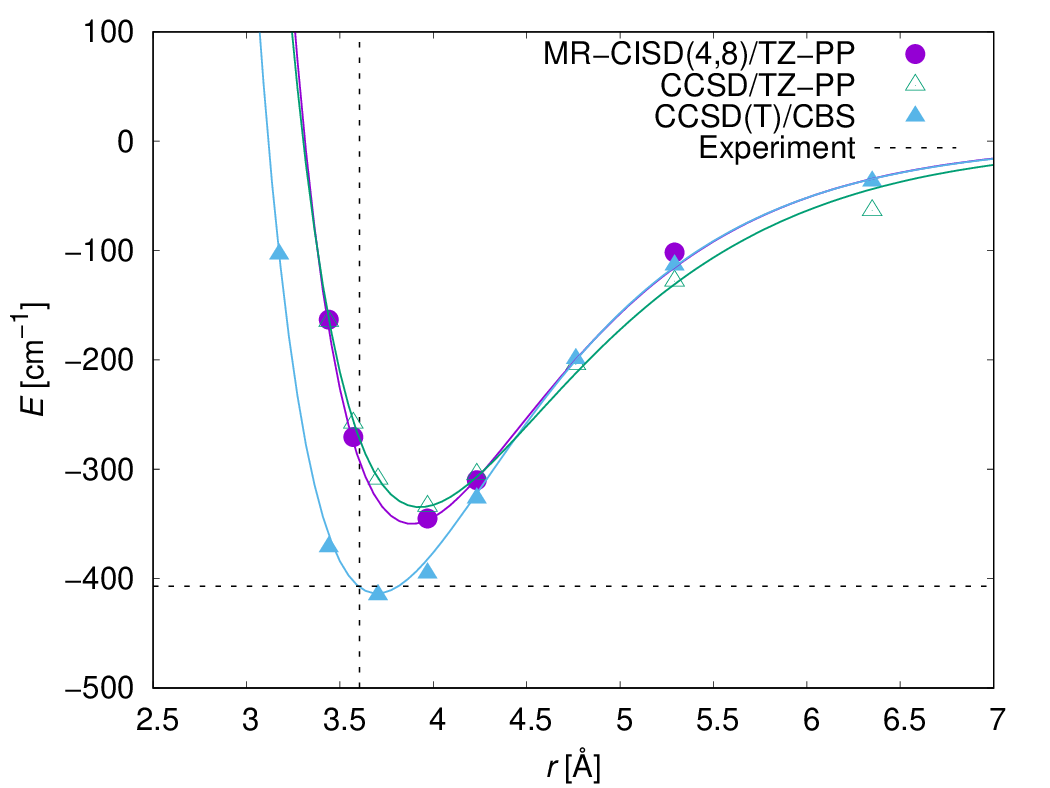}
    \caption{Potential energy curves of the ground state of the mercury dimer. The CCSD(T) complete basis set limit was extrapolated from TZ, QZ, and 5Z calculations. The MR-CISD ground state was calculated as an average from each separate irreducible representation excited state calculation.}
    \label{fig:Hg2_gs}
\end{figure}

Before discussing the excited states, we need to discuss the underlying ground state, which serves as the reference for the construction of the excited state PEC. Instead of constructing the PEC using an independent high-accuracy method such as CCSD(T), the ground states obtained within the excited-state calculations were used, all of which are depicted in Fig.~\ref{fig:Hg2_gs}.

The ground state equilibrium inter-atomic distance and dissociation energy of the mercury dimer have been rigorously examined by experimentalists~\cite{Koperski1994, Koperski1997, Koperski2008, vanZee1988, Greif2000}, and theoreticians alike~\cite{Schwerdtfeger2001, Gaston2006, Peterson2005, Luke2006, Pahl2010, Paulus2004} without reaching a definitive consensus on the values. From the most recent works the equilibrium distance is reported to be 3.605~\AA{} from experimental observations of Koperski et al.~\cite{Koperski2008} and 3.679~\AA{} from CBS/CCSD(T) with SO and full triple correction ($\Delta$T) by Pahl et al.~\cite{Pahl2010} (for pure CBS/CCSD(T) the reported value was 3.686).
The most accurate experimental value of the dissociation energy was reported by Greif~\cite{Greif2000} to be 407 cm$^{-1}$ with theory~\cite{Pahl2010} predicting 392 cm$^{-1}$ at the CBS/CCSD(T)+SO+$\Delta$T level (405 cm$^{-1}$ from CBS/CCSD(T)). From their analysis, Pahl et al.~\cite{Pahl2010} concluded that the true dissociation energy should lie lower than most experimental observations around 400 and 425 cm$^{-1}$. For our purposes, we will be comparing our results with those of Greif~\cite{Greif2000} for the dissociation energy (407 cm$^{-1}$) and Koperski et al.~\cite{Koperski2008} for the equilibrium distance (3.605~\AA{}).

From our calculations, the obtained ground states are similar in character, with the calculated dissociation energies differing from the experimental value by only 175 cm$^{-1}$. The MR-CISD (-349 cm$^{-1}$) and CCSD (-334 cm$^{-1}$) calculations predict very similar and slightly smaller dissociation energy than the experimental value (-407 cm$^{-1}$), and CCSD(T) (-413 cm$^{-1}$) slightly overestimates the binding energy. The major difference is in the predicted equilibrium distance, obtained from a Morse potential fit. CBS CCSD(T) (3.698~\AA{}) agrees well with the previous theoretical calculations but is still $\approx$0.1~\AA{} off from the experimental value of 3.605~\AA{}. The CCSD and MR-CISD approaches predict equilibrium distances of 3.927~\AA{} and 3.884~\AA{}, which deviate even further from the experiment. Since excitation energies are two orders of magnitude larger than differences in binding energy, their impact on the excited-state PEC will be almost negligible. In contrast, the equilibrium bond length has a significant effect on the spectrum. We chose the minima of the respective methods for spectrum simulation rather than experimental/CCSD(T) because they should correspond better to experimental observations of vertical excitations and better reproduce the vibrational structure of the lines if one wishes to derive it from the obtained PEC.

\subsubsection{Asymptote and size consistency}
\begin{figure}[ht]
    \centering
    \includegraphics[width=0.6\linewidth]{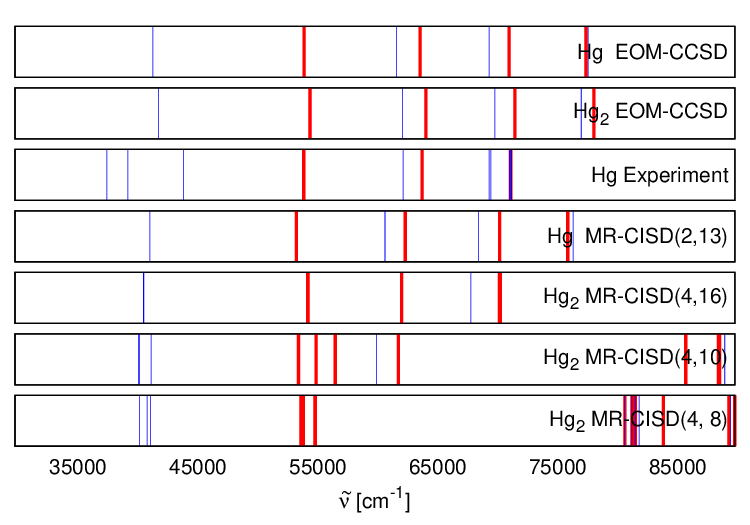}
    \caption{Energy levels of excited states of mercury atom and mercury dimer in the dissociation limit. Triplet states are coloured blue and singlets in red. Excited states were calculated using aug-cc-PVTZ-PP basis sets and a frozen core of 68 electrons, with EOM-CCSD and MR-CISD(2,13) for the atom and MR-CISD(4,8) and (4,16) for the dimer.}
    \label{fig:Hg2_asym}
\end{figure}

For the assessment of the accuracy of the employed methods in the dimer, we have two sets of experimental data with which we can compare our results. The first is the asymptotic dissociation limit, for which we have a direct experimental comparison in the form of atomic energy levels. Our results are shown in Fig.~\ref{fig:Hg2_asym}.

EOM-CCSD shows minimal difference from its atomic equivalent, apart from a slight blue shift of the whole spectrum ($\approx$ 200 cm$^{-1}$) and a small split between the $^1$D and $^3$D states. Although EOM-CCSD is formally size-consistent, the employed implementation uses a preliminary CIS calculation to generate the initial guess for the states, which is not size-consistent; this may explain the observed small differences. For the MR-CISD calculation, we first had to find an adequate active space size to capture all important transitions. While on average, MR-CISD(4,8)captures the lowest triplet and singlet P states, it shows significant non-degeneracy of these excitations as compared to EOM-CC, and more importantly, it fails to properly describe all higher excitations. MR-CISD(4,16) produces a reasonable order and energy of all excitations up to the problematic D states. Size-consistency errors in MR-CISD(4,16) are more pronounced than in EOM-CCSD; particularly non-degeneracies and state-dependent energy shifts, in contrast to the nearly uniform blue shift observed in EOM-CCSD.

We will now contrast our results with previous theoretical works. Only Balasubramanian et al. (CAS-SCF with custom basis set)~\cite{Balasubramanian1992} and Czuchaj et al. (MR-CISD with custom basis set)~\cite{Czuchaj1997}  report results for a separate mercury atom calculation and only Czuchaj et al. explicitly state that Hg$_2$ asymptotic excitation energies were comparable with the sum of atomic energies for a given state. Both atomic and asymptotic values agree well with experiment (red-shifted by approx. 2500 cm$^ {-1}$). This demonstrates the robustness of their approach given the computational limitations at the time. This methodology is irreproducible due to a custom-built basis set as well as its selective choice of reference configurations for the MR-CISD calculations. Together with their choice of a frozen core of 78 electrons, which we have shown leads to systematic deviations, this leads us to conclude that the results, while agreeing with the experiment, may partially result from the compensation of errors.

The other theoretical works of Kullie and Krosnicki do not give explicit mercury atom excited state energies, so only their asymptotic values can be evaluated. Kullie shows asymptotic values for the first six atomic excited states calculated by a relativistic spin-free Hamiltonian; unfortunately, the work does not state the energy of the ground state calculated by the Dirac-Coulomb Hamiltonian, and therefore, the asymptotes with the inclusion of SOC can not be analysed. The spin-free values are in relatively good agreement with experiment, although on average worse than those of Czuchaj et al. For the first two asymptotes $^{3}$P (6s6p) and $^{1}$P (6s6p), the values are red shifted by $\approx$ 5000 cm$^{-1}$ while for higher excited states, they report a blue shift of $\approx$ 2000 cm$^{-1}$.\\

Krosnicki et al. concentrate only on the first few excited states corresponding to the $^3$P$_{1}$ asymptote. Krosnicki et al. asymptotic values are identical with atomic excitation energies in the NIST database, leading us to conclude that the curves were calculated separately and shifted to fit experimental values. In regard to works where the PEC was obtained as a whole (Czuchaj et al. and Kullie) rather than constructed and shifted to experimental values (Krosnicki), we managed to obtain asymptotic values in better agreement with experiment, with a reproducible methodology and a rigorous treatment of electron correlation.\\

\subsubsection{Potential energy curve}
\label{sec:Hg2_PES}
\begin{figure}[ht]
    \centering
    \includegraphics[width=0.6\linewidth]{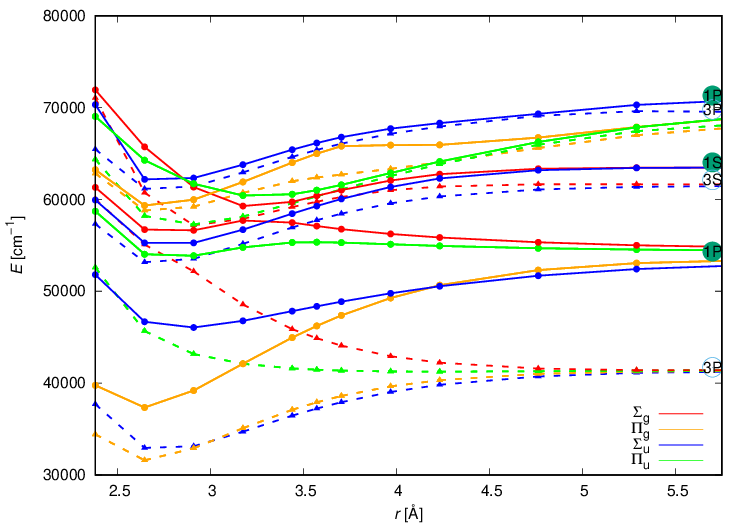}
    \caption{Potential energy curve of excited states of the mercury dimer calculated with EOM-CCSD/aug-cc-PVTZ-PP constructed above the CCSD ground state reference. Singlet states are represented by full lines, while triplet states use dashed lines. Zero energy was set to the dissociation limit of the mercury dimer.}
    \label{fig:Hg2_PES}
\end{figure}

The EOM-CCSD potential energy curve of the mercury dimer is depicted in Fig.~\ref{fig:Hg2_PES} with the transition dipole moments of the excited states shown in Fig.~S4. 
This PEC was constructed using the CCSD ground state shown in Fig.~\ref{fig:Hg2_gs}. Since we have observed that the inclusion of perturbative triple excitations shifted the equilibrium distance and, to a significantly smaller degree, the dissociation energy in the ground state, similar shifts are expected for the excited states. However, in the case of excitation energies, the $E_{\mathrm{CCSD}}-E_{\mathrm{CCSD(T)}}$ errors should cancel out, as excitation energies are energy differences. This "correction" is compounded by the fact that the differences in excitation energies between two interatomic distances are two orders of magnitude larger than the $E_{\mathrm{CCSD}}$-$E_{\mathrm{CCSD(T)}}$ errors and therefore Fig.~\ref{fig:Hg2_PES} represents the most accurate depiction of the electronic states of the mercury dimer to date.
%E_disoc CCSD (334) CCSDT (416)  diff  (82)   cm-1 
%1st E_exc S  3.6 (49601)  3.9 (50963) diff (1362) cm-1

From the calculated PEC, we were able to extract spectroscopic parameters and compare them with the experimental observations of mercury dimer absorption spectra~\cite{Koperski1994, Koperski1994Can, Koperski1997, Koperski2008, Schlauf1989}. Most experimental observations employed fluorescence spectroscopy and therefore observed ground $^{1}\mathrm{\Sigma_{g}}^{+}$ (X0$^{+}_{\mathrm{g}}$) to triplet state transitions, with only one singlet $\rightarrow$ singlet transition reported by Koperski et al.~\cite{Koperski2008}. A comparison of the calculated and experimental transitions is compiled in Tab.~\ref{tab:Hg2_state}. 
\begin{table}[ht]
\caption{Spectroscopic parameters of the ground and excited states of mercury dimer. Theoretical values were calculated using CCSD for the ground state and EOM-CCSD for excited states with aug-cc-PVTZ-PP basis set with 68 frozen core electrons.}
\begin{tabular}{|ll|ll|ll|ll|}
\hline
\multicolumn{2}{|c|}{State}        & \multicolumn{2}{c|}{$r_{\mathrm{eq}}$ [\AA{}]} & \multicolumn{2}{c|}{$E_{\mathrm{dissoc}}$ [cm$^{-1}$]} & \multicolumn{2}{c|}{$\nu_{0\rightarrow0}$ [cm$^{-1}$]} \\
Theor.        & Exp. & Theor.        & Exp.         & Theor.          & Exp.          & Theor.        & Exp.         \\
\hline
$^{1}\mathrm{\Sigma_{g}}^{+}$ & X0$^{+}_{g}$~\cite{Schlauf1989}   & 3.895       & 3.63       & 334           & 360          & -           & -           \\
$^{3}\mathrm{\Sigma_{u}}^{+}$ & 1$_{u}$~\cite{Koperski1994Can} & 2.646       & 2.50        & 8544          & 8100         & 32947       & 31340       \\
$^{1}\mathrm{\Sigma_{u}}^{+}$ & 0$^{+}_{u}$~\cite{Koperski2008}   & 2.910        & 3.00 & 8041          & 8320         & 46062       & 46201    \\  
\hline
\end{tabular}
\label{tab:Hg2_state}
\end{table}
The ground state was discussed in its own section and is mentioned only for completeness. Experimentally several singlet $\rightarrow$ triplet 1$_{u}$~\citep{Koperski1994Can,Koperski2008} transitions were observed which correspond to $^{3}\mathrm{\Pi_{u/g}}$ states on our PEC. These states experience spin-orbit coupling and therefore cannot be directly compared with the present results. 
The $^{3}\mathrm{\Sigma_{u}}^{+}$  (1$_{u}$) state is not affected by SOC, allowing direct comparison with experiment. We report good agreement in both the equilibrium distance and dissociation energy ($\Delta E$= 466 cm$^{-1}$). The $^{1}\mathrm{\Sigma_{u}}^{+}$ (0$^{+}_{u}$) is most important for our study since it is the only experimentally reported excited singlet state of mercury. For this state, the calculated values of equilibrium distance and dissociation energy are in excellent agreement with experiment. We did not perform a vibrational analysis of our PEC; we therefore associate the $\nu_{0\rightarrow0}$ transitions with the minima of our curves. Even with this approximation, the calculated values are close to those measured in experiments in case of $^{3}\mathrm{\Sigma_{u}}^{+}$ $\Delta E$= 1607 cm$^{-1}$ and $^{1}\mathrm{\Sigma_{u}}^{+}$ $\Delta E$= 139 cm$^{-1}$. The excellent agreement between the experimental and calculated spectroscopic constants of the $^{1}\mathrm{\Sigma_{u}}^{+}$  excited state is key for the validation of our methodology.

\subsection{Small clusters}

We performed full and constrained scans of the ground-state structural potential energy surfaces of Hg$_{3}$-Hg$_{5}$ to confirm the existence of local minima other than the global high-symmetry ones. The only local minima found was for the linear Hg$\mathrm{_{3}}$ structure with both bond distances equal to 3.89 \AA{} and a well depth of only 23 cm$^{-1}$. Due to this very weakly bound state, we do not expect it to have any noticeable effects on the resulting absorption spectra. For the mercury tetramer and pentamer, we have not detected any local minima. We discuss this analysis in detail in the Supplementary Material, and we continue with global minima later in the text.

\begin{table}[bp!]
\caption{Equilibrium nearest-neighbour distance $r_{\mathrm{eq}}$, dissociation energy $E_{\mathrm{dissoc}}$ and HOMO-LUMO gap for Hg$_{\mathrm{1-5}}$ calculated at various levels of theory.}
\begin{tabular}{|l|cc|cc|c|}
\hline
N & \multicolumn{2}{c|}{$r_{\mathrm{eq}}$ [\AA]} &  \multicolumn{2}{c|}{$E_{\mathrm{dissoc}}$ [cm$^{-1}$]} & $E_{\mathrm{HOMO-LUMO}}$  [cm$^{-1}$]  \\
  & CCSD           & CCSD(T)                                 & CCSD             & CCSD(T)     & MP2        \\
  \hline
1 & -     & -     & -    & -    & 76641 \\
2 & 3.895 & 3.699 & 334  & 482  & 69974 \\
3 & 3.810 & 3.612 & 1154 & 1787 & 68768 \\
4 & 3.673 & 3.459 & 1151 & 2143 & 70041 \\
5 & 3.664 & 3.504 & 2587 & 4876 & 64190 \\           
\hline
\end{tabular}
    \label{tab:Hg_clusters}
\end{table}

We present calculated equilibrium nearest-neighbour distances, dissociation energies, and HOMO-LUMO gaps for Hg$_{\mathrm{1-5}}$ ground states in Tab.~\ref{tab:Hg_clusters}. The geometries considered were an equilateral triangle (Hg$_3$), a tetrahedron (Hg$_4$), and a trigonal bipyramid (Hg$_5$). In Tab.~\ref{tab:Hg_clusters}, r$_{eq}$ represents the optimized edge length of the polyhedra as well as the nearest-neighbour bond distances. As with the dimer, we see that CCSD systematically overestimates the bond lengths and underestimates the dissociation energies, with a sharp increase in CCSD/CCSD(T) error for Hg$_{\mathrm{4}}$ and Hg$_{\mathrm{5}}$. This is not surprising since mercury clusters are weakly bound, mainly through van der Waals forces, and therefore even small losses in correlation energy accumulate with increasing cluster size.

Compared with similar studies on other elements (e.g., Be)~\citep{Labanc2018}, the dissociation energy increases considerably more slowly than in Be, emphasizing the weakly bound character of these clusters. The PEC of the ground states for Hg$_{\mathrm{3-5}}$ can be found in Fig.~S11. However, the arguments made for the excitation energies of the dimer in section~\ref{sec:Hg2_PES} are valid for larger clusters as well, and calculated absorption spectra are expected to remain valid. Paradoxically, the difference in CCSD/CCSD(T) bond lengths decreases with cluster size, although for any general conclusions on the reason why, larger cluster sizes would need to be examined.

For cluster sizes larger than 2, the computational costs to perform MR-CISD become unreasonable, and only EOM-CCSD remains feasible. Though direct comparison of cluster properties is no longer possible, there are still several parameters that can be used to assess the multi-reference character of the ground state. One of these is the HOMO-LUMO gap shown in Tab.~\ref{tab:Hg_clusters}. In all studied clusters, the HOMO-LUMO gap is quite large $\approx$ 65000-75000 cm$^{-1}$, indicating a low probability of mixing between the ground and lowest excited states. The large and slowly decreasing gap further indicates that these clusters remain weakly bonded, which is supported by previous studies where covalent character emerges for cluster sizes larger than approximately 10~\cite{Rademann1992}. Another parameter that was examined was the T$_{1}$ diagnostic~\cite{Jiang2012} from the CCSD/CCSD(T) calculations. For all equilibrium geometries (including atom and dimer), the T$_{1}$ remained below 0.015 (Fig.~S12), indicating that a single-reference description is adequate.~\cite{Wang2015, Coe2015} Another parameter that was calculated was the energy difference between RHF and UHF, which indicates spin symmetry breaking. The $E_\mathrm{RHF}-E_\mathrm{UHF}$ differences were zero for all equilibrium geometries in all clusters,  supporting the single-reference character~\citep{Staroverov2000}.

\subsection{Absorption spectra}
\begin{figure*}[ht]
    \centering
    \includegraphics[width=0.32\linewidth]{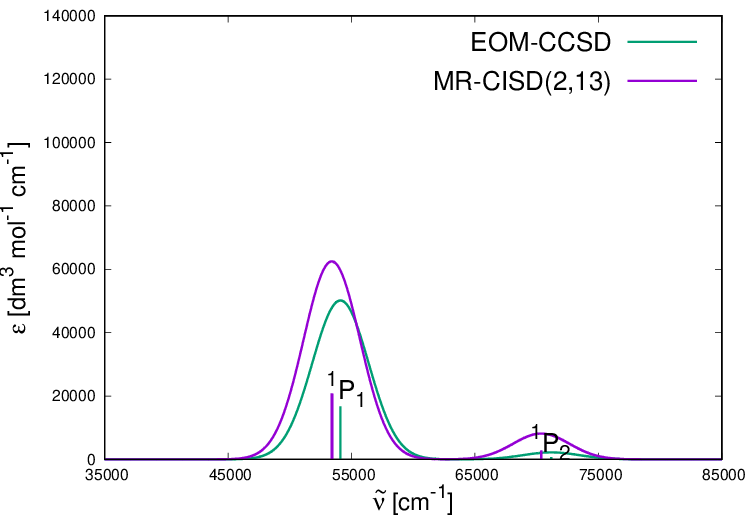}
    \includegraphics[width=0.32\linewidth]{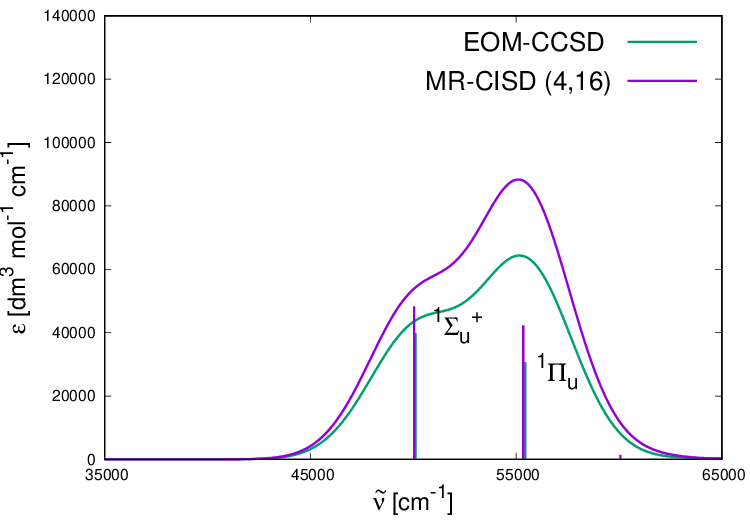}
    \includegraphics[width=0.32\linewidth]{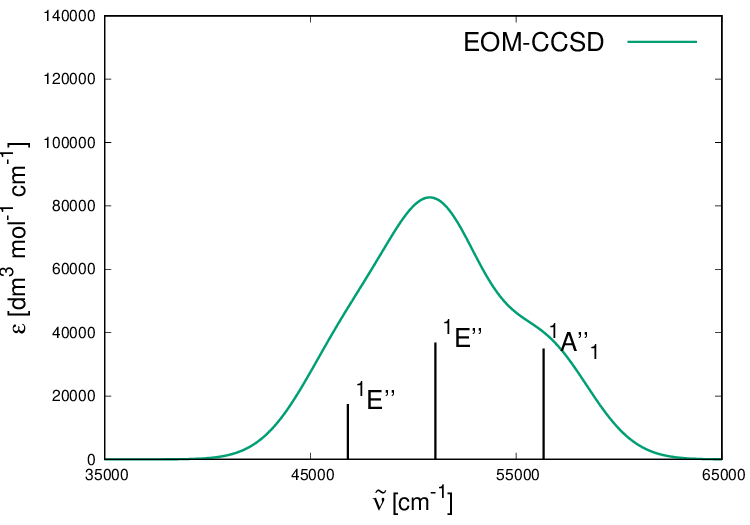}
    \includegraphics[width=0.32\linewidth]{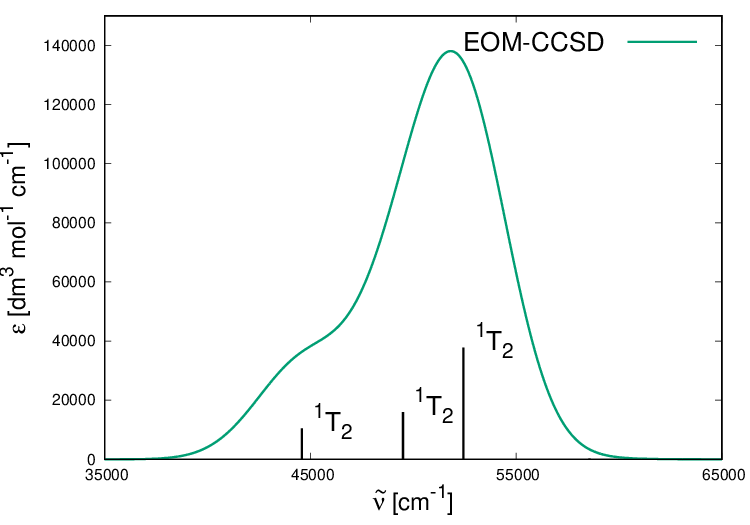}
    \includegraphics[width=0.32\linewidth]{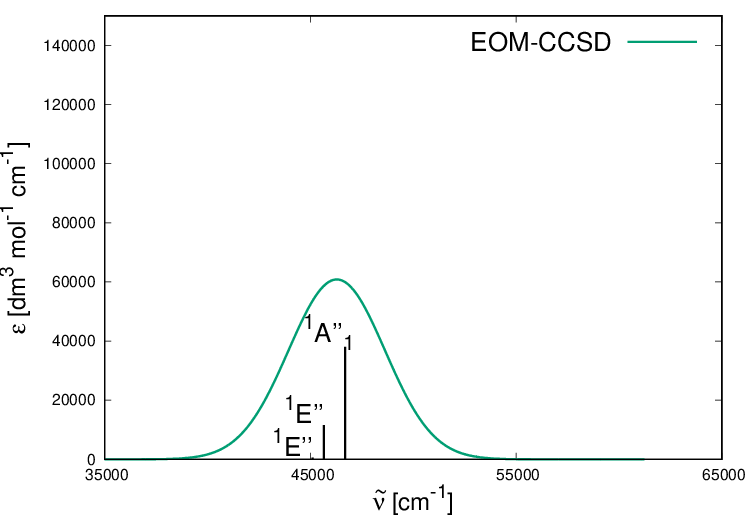}
    \caption{Simulated absorption spectrum of Hg$_{\mathrm{1-5}}$ in units of the molar absorption coefficient. Excited states were calculated with EOM-CCSD/aug-cc-PVTZ-PP basis sets and frozen core of 68 electrons. For the absorption spectrum of the mercury atom and dimer, we also show the spectrum obtained by MR-CISD(2,13)/aug-cc-PVTZ-PP. The spectra include the particular optically active excited states together with their respective irreducible representations.}
    \label{fig:Hg_spectrum}
\end{figure*}

The calculated absorption spectra of Hg$_{\mathrm{1-5}}$ are depicted in Fig.~\ref{fig:Hg_spectrum}. The spectra were simulated using excitation energies and oscillator strengths obtained by EOM-CCSD with aug-cc-PVTZ-PP basis sets and a frozen core of 68 electrons, with cluster geometries optimized at the CCSD level (r$_{\mathrm{eq}}$ from Tab.~\ref{tab:Hg_clusters}). 

The oscillator strengths are obtained from the transition dipole moments between the electronic ground and excited states at the corresponding ground-state equilibrium geometry. Thus, the calculated absorption spectra represent vertical electronic transitions from the ground-state minimum, with the oscillator strengths evaluated at this single geometry. The absorption spectrum was approximated by applying Gaussian broadening to these discrete transitions.

In the case of the mercury atom and dimer, we also show the MR-CISD absorption spectrum. The spectra were simulated as a sum of bands, each approximated by a Gaussian function with a bandwidth of 3226 cm$^{-1}$. All considered excited states together with their oscillator strengths are compiled in Tab.~S6 and Tab.~S7.

In the atomic spectra only two transitions showed a non-zero oscillator strength: $^{1}$S$_{0}$ $\rightarrow$ $^{1}$P$_{1}$  at 54103/53446 cm$^{-1}$ and $^{1}$S$_{0}$ $\rightarrow$ $^{1}$P$_{2}$ at 71308/70385 cm$^{-1}$ with an oscillator strength ratio of 1:0.04 (EOM-CCSD/aug-cc-PVTZ-PP) and 1:0.09 (MR-CISD/aug-cc-PVTZ-PP) respectively. Our EOM-CCSD results are in excellent agreement with experimental observations of atomic absorption line spectra~\cite{NIST_ASD}, where both of these transitions (54068$^{-1}$ and 71294$^{-1}$) are reported as being optically active. The Peak ratio in the NIST database is reported to be 1:0.006, but the lines were sourced from different experiments, and therefore their relative intensities are only qualitative.

The mercury dimer shows two optically allowed transitions: $^{1}\mathrm{\Sigma_{u}}^{+}$ and a doubly degenerate $^{1}\mathrm{\Pi_{u}}$. The $^{1}\mathrm{\Sigma_{u}}^{+}$ state corresponds to transitions from HOMO to the virtual $\sigma_{2}$ and $\sigma_{3}$ orbitals and from HOMO$-1$ to the virtual $\mathrm{\sigma}^{*}_{1}$ orbital. The $^{1}\mathrm{\Pi_{u}}$ state corresponds to transitions from the HOMO to virtual $\pi_{1}$ and $\pi_{2}$ and HOMO$-1$ to $\pi^{*}_{1}$ and $\pi^{*}_{1}$ orbitals. The EOM-CCSD and MR-CISD peaks have almost identical excitation energies and very similar peak ratios (1:0.76 for EOM-CCSD and 1:0.87 for MR-CISD), to an even greater extent than for the mercury atom, further confirming the reliability of our methodology. Experimentally, only the $^{1}\mathrm{\Sigma_{u}}^{+}$ transition has been observed~\cite{Koperski2008} via laser-induced fluorescence (LIF). While LIF provides information on emissive states, it does not directly probe absorption spectra. Our excellent agreement with experimental atomic spectra further supports the accuracy of our methodology, suggesting that the predicted dimer absorption spectrum provides a reliable description of the electronic transitions. To the best of our knowledge, no experimental absorption spectra for the mercury dimer have been reported; therefore, our results provide, to our knowledge, the first theoretical prediction of the mercury dimer absorption spectrum.

The transitions in larger mercury clusters follow a similar trend to the atom and dimer, where the visible transitions are from frontier occupied orbitals, HOMO${-n}$, where n goes from 0 to the cluster size, to the lowest high symmetry virtual orbitals. All mercury clusters, except for the atom, have wide absorption peaks in approximately the same region $\approx$ 37000 to 60000 cm$^{-1}$, which is not surprising given their very similar HOMO-LUMO gaps. This overlap of absorption bands would make them somewhat difficult to distinguish in mixed samples, but on the other hand, each cluster has a characteristic spectrum shape which would clearly identify it in adequately sized and separated samples. 

%\MN{}{Rozsirenie: Vieme spocitat triplet$\rightarrow$ absorpcne triplet prechody.}
%\MN{}{Rozsirenie: EOM viem spocitat emisne spektrum. K nemu je viac experimentov. Url v tex commentari.}
%https://www.physics.nist.gov/PhysRefData/Handbook/Tables/mercurytable3.htm
%
%===================================
\section{Conclusions}
\label{conclusions}
In this work, we have systematically evaluated the performance of various basis sets and computational approaches for the study of the mercury atom and small clusters Hg$_{2-5}$. Our findings demonstrate that effective core potentials (ECPs) are sufficient to accurately describe scalar relativistic effects in the mercury atom and dimer, offering a computationally efficient alternative to more demanding all-electron approaches.

The ANO-RCC-XZ basis sets provide a reliable description of lower excited states but fail to adequately represent higher excited states due to insufficient flexibility in the high-energy virtual orbital space. Among the tested basis sets, the aug-cc-PVXZ-DKH family delivers the most accurate and comprehensive results, offering an accurate representation of high-lying virtual orbitals, incorporation of relativistic effects through the Douglas–Kroll–Hess (DKH) transformation, and the ability to perform spin–orbit coupling (SOC) calculations. On the other hand, the aug-cc-PVXZ-PP basis sets provide an excellent compromise, providing accurate descriptions of highly excited states while being significantly more computationally efficient.

For the mercury atom and dimer, we report excellent agreement of the equation-of-motion coupled-cluster (EOM-CCSD) results with multi-reference configuration interaction with singles and doubles (MR-CISD) and experimental data, confirming the reliability of EOM-CC for describing excited states in such systems as well as the single-reference character of these systems.

Using this benchmark, we employed EOM-CCSD together with aug-cc-PVTZ-PP to calculate excitation energies and oscillator strengths of the mercury atom and small clusters (Hg$_{\mathrm{1-5}}$) from which we constructed, for the first time, \textit{ab initio} absorption spectra. The multi-reference character of clusters larger than the dimer was assessed indirectly using T$_{1}$ diagnostic, HOMO-LUMO gap, and RHF/UHF energy difference. All of these parameters indicated a single-reference ground state, supporting the use of EOM-CC.

Our optimized methodology makes studying absorption spectra of even larger clusters computationally feasible while maintaining a high level of accuracy and control. The calculated spectra should provide a good theoretical reference for the identification of cluster sizes in molecular beam and low-temperature experiments of mercury systems, as well as serve as a benchmark for "cheaper" theoretical methods in the analysis of larger systems.

\section*{Supplementary Material:}
Section discussing the effect of basis sets and froze core approximation on the orbital ordering of Hg atom.
Section discussing the search of local minima in the structural space of Hg$_3$-Hg$_5$.
Energy levels of the Hg atom from various methods and setups. Irreducible representations for Hg dimer, trimer, and pentamer. 
Potential energy curves and transitional dipole moment curves of Hg dimer at various levels of methodology.
Tabulated excited states of Hg$_{1-5}$ calculated by EOM-CCSD.
Tabulated excited states of Hg atom and dimer calculated by MR-CISD.
Potential energy curves of Hg$_{3-5}$ ground states. 
T1 diagnostic for the ground states of Hg$_{2-5}$ clusters.

\section*{Acknowledgments}
We thank René Kalus for initial discussions.
This article has been produced with the financial support of the European Union under the LERCO project number CZ.10.03.01/00/22\_003/0000003 via the Operational Programme Just Transition. The computations were performed at IT4Innovations National Supercomputing Center through the e-INFRA CZ (ID:90140).

\section*{Data Availability Statement}
The data underlying this study are available in the published article and its Supporting Information.

%((Additional Supporting Information may be found in the online version of this article.))
\clearpage
%%%%%%%%%%%%%%%%%%%%%%%%%%%%%%%%%%%%%%%%%%%%%%%%%%%%%%%%%%%%%%%%%%%%%%%%%%%%%%%%%
% BIBLIOGRAPHY
\bibliography{bibliography}  

\begin{thebibliography}{10}

\bibitem{Brechignac1985}
Brechignac, C.; Broyer, M.; Cahuzac, P.; Delacretaz, G.; Labastie, P.;
  W\"{o}ste, L. Size dependence of inner-shell autoionization lines in mercury
  clusters {\em Chemical Physics Letters} {\bf 1985}, {\em 120}(6), 559--563.

\bibitem{Brchignac1988}
Br\'{e}chignac, C.; Broyer, M.; Cahuzac, P.; Delacretaz, G.; Labastie, P.;
  Wolf, J.~P.; W\"{o}ste, L. Probing the transition from van der waals to
  metallic mercury clusters {\em Physical Review Letters} {\bf 1988}, {\em
  60}(4), 275--278.

\bibitem{Pastor1988}
Pastor, G.~M.; Stampfli, P.; Bennemann, K.~H. On the transition from van der
  waals- to metallic bonding in hg-clusters as a function of cluster size {\em
  Physica Scripta} {\bf 1988}, {\em 38}(4), 623--626.

\bibitem{Kitamura2006}
Kitamura, H. Theoretical potential energy surfaces for excited mercury trimers
  {\em Chemical Physics} {\bf 2006}, {\em 325}(2), 207--219.

\bibitem{Garcia1991}
Garcia, M.~E.; Pastor, G.~M.; Bennemann, K.~H. Theory for the change of the
  bond character in divalent-metal clusters {\em Physical Review Letters} {\bf
  1991}, {\em 67}(9), 1142--1145.

\bibitem{Rademann1987}
Rademann, K.; Kaiser, B.; Even, U.; Hensel, F. Size dependence of the gradual
  transition to metallic properties in isolated mercury clusters {\em Physical
  Review Letters} {\bf 1987}, {\em 59}(20), 2319--2321.

\bibitem{Harbour1971}
Harbour, P.~J. The detection and ionization potential of mercury clusters {\em
  Journal of Physics B: Atomic and Molecular Physics} {\bf 1971}, {\em 4}(4),
  528--530.

\bibitem{BCabaud1980}
Cabaud, B.; Hoareau, A.; Melinon, P. Time-of-flight spectroscopy of a
  supersonic beam of mercury. intensities and appearance potentials of
  hgxaggregates {\em Journal of Physics D: Applied Physics} {\bf 1980}, {\em
  13}(10), 1831--1834.

\bibitem{Yannouleas1990}
Yannouleas, C.; Pacheco, J.~M.; Broglia, R.~A. Microscopic structure of the
  plasma resonance in charged potassium microclusters {\em Physical Review B}
  {\bf 1990}, {\em 41}(9), 6088--6091.

\bibitem{Fallgren1990}
Fallgren, H.; Martin, T. Photoabsorption of cs8 and cs10o clusters {\em
  Chemical Physics Letters} {\bf 1990}, {\em 168}(3), 233--238.

\bibitem{Rademann1992}
Rademann, K.; Dimopoulou-Rademann, O.; Schlauf, M.; Even, U.; Hensel, F.
  Evolution of surface plasmon resonance absorption in large gas phase clusters
  of mercury: Approaching the bulk {\em Physical Review Letters} {\bf 1992},
  {\em 69}(22), 3208--3211.

\bibitem{Kitamura20062}
Kitamura, H. Analysis of excited mercury clusters with diatomic potential
  energy curves {\em Chemical Physics Letters} {\bf 2006}, {\em 425}(4),
  205--209.

\bibitem{Vitek2022}
V\'{\i}tek, A.; Kalus, R. Thermodynamics of small mercury clusters and the role
  of electronically excited states: a case study on {Hg$_{13}$} {\em Phys.
  Chem. Chem. Phys.} {\bf 2022}, {\em 24}, 6915--6925.

\bibitem{Sarmanova2019}
\v{S}armanov\'{a}, M.; V\'{\i}tek, A.; \'{C}osi\'{c}, R.; Kalus, R.
  Photoabsorption markers of pressure-induced phase changes in small mercury
  clusters. a case study on {Hg$_8$} {\em RSC Adv.} {\bf 2019}, {\em 9},
  37258--37266.

\bibitem{Czuchaj1997}
Czuchaj, E.; Rebentrost, F.; Stoll, H.; Preuss, H. Calculation of ground- and
  excited-state potential energy curves for the {Hg$_2$} molecule in a
  pseudopotential approach {\em Chemical Physics} {\bf 1997}, {\em 214}(2),
  277--289.

\bibitem{Kullie2014}
Kullie, O. A relativistic time-dependent density functional study of the
  excited states of the mercury dimer {\em The Journal of Chemical Physics}
  {\bf 2014}, {\em 140}(2), 024304.

\bibitem{Kronicki2015}
Kro\'{s}nicki, M.; Strojecki, M.; Urba\'{n}czyk, T.; Pashov, A.; Koperski, J.
  Interatomic potentials of the heavy van der waals dimer {Hg$_2$}: A
  "test-bed" for theory-to-experiment agreement {\em Physics Reports} {\bf
  2015}, {\em 591}, 1--31.

\bibitem{Balasubramanian1992}
Balasubramanian, K.; Das, K.; Liao, D. Spectroscopic constants and potential
  energy curves for {Hg$_2$} {\em Chemical Physics Letters} {\bf 1992}, {\em
  195}(5), 487--493.

\bibitem{Calvo2012}
Calvo, F.; Pahl, E.; Schwerdtfeger, P.; Spiegelman, F. Diatomics-in-molecules
  modeling of many-body effects on the structure and thermodynamics of mercury
  clusters {\em Journal of Chemical Theory and Computation} {\bf 2012}, {\em
  8}(2), 639--648.

\bibitem{Cosi2018}
\'{C}osi\'{c}, R.; V\'{\i}tek, A.; Kalus, R. Photoabsorption spectra of small
  mercury clusters: a computational study {\em Physical Chemistry Chemical
  Physics} {\bf 2018}, {\em 20}(45), 28871--28880.

\bibitem{Pritchard2019}
Pritchard, B.~P.; Altarawy, D.; Didier, B.; Gibson, T.~D.; Windus, T.~L. New
  basis set exchange: An open, up-to-date resource for the molecular sciences
  community {\em Journal of Chemical Information and Modeling} {\bf 2019}, {\em
  59}(11), 4814--4820.

\bibitem{RN204}
Neese, F. Software update: The orca program system--version 5.0 {\em WIREs
  Computational Molecular Science} {\bf 2022}, {\em 12}(5), e1606.

\bibitem{RN155}
Kollmar, C.; Sivalingam, K.; Helmich-Paris, B.; Angeli, C.; Neese, F. A
  perturbation-based super-ci approach for the orbital optimization of a casscf
  wave function {\em J. Comput. Chem.} {\bf 2019}, {\em 40}, 1463--1470.

\bibitem{RN205}
Neese, F. The shark integral generation and digestion system {\em J. Comp.
  Chem.} {\bf 2023}, {\em 44}(3), 381--396.

\bibitem{RN229}
Ugandi, M.; Roemelt, M. A recursive formulation of one-electron coupling
  coefficients for spin-adapted configuration interaction calculations
  featuring many unpaired electrons {\em Int. J. Quantum Chem.} {\bf 2023},
  {\em 123}, e27045.

\bibitem{RN84}
Neese, F. The orca program system {\em WIRES Comput. Molec. Sci.} {\bf 2012},
  {\em 2}(1), 73--78.

\bibitem{RN139}
Neese, F. Software update: the orca program system, version 4.0 {\em WIRES
  Comput. Molec. Sci.} {\bf 2018}, {\em 8}(1), 1--6.

\bibitem{RN178}
Neese, F.; Wennmohs, F.; Becker, U.; Riplinger, C. The orca quantum chemistry
  program package {\em J. Chem. Phys.} {\bf 2020}, {\em 152}, L224108.

\bibitem{RN21}
Neese, F. Approximate second-order scf convergence for spin unrestricted
  wavefunctions {\em Chem. Phys. Lett.} {\bf 2000}, {\em 325}(1), 93--98.

\bibitem{g16}
Gaussian 16 {R}evision {C}.01. Frisch, M.~J.; Trucks, G.~W.; Schlegel, H.~B.;
  Scuseria, G.~E.; Robb, M.~A.; Cheeseman, J.~R.; Scalmani, G.; Barone, V.;
  Petersson, G.~A.; Nakatsuji, H.; Li, X.; Caricato, M.; Marenich, A.~V.;
  Bloino, J.; Janesko, B.~G.; Gomperts, R.; Mennucci, B.; Hratchian, H.~P.;
  Ortiz, J.~V.; Izmaylov, A.~F.; Sonnenberg, J.~L.; Williams-Young, D.; Ding,
  F.; Lipparini, F.; Egidi, F.; Goings, J.; Peng, B.; Petrone, A.; Henderson,
  T.; Ranasinghe, D.; Zakrzewski, V.~G.; Gao, J.; Rega, N.; Zheng, G.; Liang,
  W.; Hada, M.; Ehara, M.; Toyota, K.; Fukuda, R.; Hasegawa, J.; Ishida, M.;
  Nakajima, T.; Honda, Y.; Kitao, O.; Nakai, H.; Vreven, T.; Throssell, K.;
  Montgomery, {Jr.}, J.~A.; Peralta, J.~E.; Ogliaro, F.; Bearpark, M.~J.; Heyd,
  J.~J.; Brothers, E.~N.; Kudin, K.~N.; Staroverov, V.~N.; Keith, T.~A.;
  Kobayashi, R.; Normand, J.; Raghavachari, K.; Rendell, A.~P.; Burant, J.~C.;
  Iyengar, S.~S.; Tomasi, J.; Cossi, M.; Millam, J.~M.; Klene, M.; Adamo, C.;
  Cammi, R.; Ochterski, J.~W.; Martin, R.~L.; Morokuma, K.; Farkas, O.;
  Foresman, J.~B.; Fox, D.~J. {\bf 2016}.

\bibitem{Visscher1997}
Visscher, L.; Dyall, K. Dirac-fock atomic electronic structure calculations
  using different nuclear charge distributions {\em Atomic Data and Nuclear
  Data Tables} {\bf 1997}, {\em 67}(2), 207--224.

\bibitem{Srivastava1992}
Srivastava, R.; Zuo, T.; McEachran, R.~P.; Stauffer, A.~D. Relativistic
  distorted-wave calculation of electron excitation of mercury {\em Journal of
  Physics B: Atomic, Molecular and Optical Physics} {\bf 1992}, {\em 25}(10),
  2409--2425.

\bibitem{Gleichmann1994}
Gleichmann, M.~M.; He\ensuremath{\beta}, B.~A. Relativistic all-electron ab
  initio calculations for the ground and excited states of the mercury atom
  {\em Chemical Physics Letters} {\bf 1994}, {\em 227}(1), 229--234.

\bibitem{Mosyagin2000}
Mosyagin, N.~S.; Eliav, E.; Titov, A.~V.; Kaldor, U. Comparison of relativistic
  effective core potential and all-electron dirac-coulomb calculations of
  mercury transition energies by the relativistic coupled-cluster method {\em
  Journal of Physics B: Atomic, Molecular and Optical Physics} {\bf 2000}, {\em
  33}(4), 667--676.

\bibitem{Koperski1994}
Koperski, J.; Atkinson, J.; Krause, L. The {0$^+_u$(6 $^3$P$_1$)X0$^+_g$}
  spectrum of {Hg$_2$} excited in a supersonic jet {\em Chemical Physics
  Letters} {\bf 1994}, {\em 219}(3), 161--168.

\bibitem{Koperski1997}
Koperski, J.; Atkinson, J.; Krause, L. The {G0$^+_u$(6$^1$P$_1$)-X0$^+_g$}
  excitation and fluorescence spectra of {Hg$_2$} excited in a supersonic jet
  {\em Journal of Molecular Spectroscopy} {\bf 1997}, {\em 184}(2), 300--308.

\bibitem{Koperski2008}
Koperski, J.; Qu, X.; Meng, H.; Kenefick, R.; Fry, E. Rotational analysis of
  the (57, 0) band of the triplet-singlet transition in {Hg$_2$} produced in a
  free-jet expansion beam {\em Chemical Physics} {\bf 2008}, {\em 348}(1),
  103--112.

\bibitem{vanZee1988}
van Zee, R.~D.; Blankespoor, S.~C.; Zwier, T.~S. Direct spectroscopic
  determination of the {Hg$_2$} bond length and an analysis of the {2540 \AA{}}
  band {\em The Journal of Chemical Physics} {\bf 1988}, {\em 88}(8),
  4650--4654.

\bibitem{Greif2000}
Greif-W\"{u}stenbecker, J.~N. The line shape of the low temperature raman
  spectrum of mercury diatoms {\em Ph.D. thesis, Philipps University} {\bf
  2000}.

\bibitem{Schwerdtfeger2001}
Schwerdtfeger, P.; Wesendrup, R.; Moyano, G.~E.; Sadlej, A.~J.; Greif, J.;
  Hensel, F. The potential energy curve and dipole polarizability tensor of
  mercury dimer {\em The Journal of Chemical Physics} {\bf 2001}, {\em
  115}(16), 7401--7412.

\bibitem{Gaston2006}
Gaston, N.; Schwerdtfeger, P. From the van der waals dimer to the solid state
  of mercury with relativistic ab initio and density functional theory {\em
  Phys. Rev. B} {\bf 2006}, {\em 74}, 024105.

\bibitem{Peterson2005}
Peterson, K.~A.; Puzzarini, C. Systematically convergent basis sets for
  transition metals. ii. pseudopotential-based correlation consistent basis
  sets for the group 11 {(Cu, Ag, Au)} and 12 {(Zn, Cd, Hg)} elements {\em
  Theoretical Chemistry Accounts} {\bf 2005}, {\em 114}(4), 283--296.

\bibitem{Luke2006}
Luke\v{s}, V.; Il\v{c}in, M.; Laurinc, V.; Biskupi\v{c}, S. On the structure
  and physical origin of van der waals interaction in zinc, cadmium and mercury
  dimers {\em Chemical Physics Letters} {\bf 2006}, {\em 424}(1), 199--203.

\bibitem{Pahl2010}
Pahl, E.; Figgen, D.; Thierfelder, C.; Peterson, K.~A.; Calvo, F.;
  Schwerdtfeger, P. A highly accurate potential energy curve for the mercury
  dimer {\em The Journal of Chemical Physics} {\bf 2010}, {\em 132}(11),
  114301.

\bibitem{Paulus2004}
Paulus, B.; Rosciszewski, K.; Gaston, N.; Schwerdtfeger, P.; Stoll, H.
  Convergence of the ab initio many-body expansion for the cohesive energy of
  solid mercury {\em Phys. Rev. B} {\bf 2004}, {\em 70}, 165106.

\bibitem{Koperski1994Can}
Koperski, J.; Atkinson, J.~B.; Krause, L. Molecular beam spectroscopy of the
  and transitions in {Hg$_2$} {\em Canadian Journal of Physics} {\bf 1994},
  {\em 72}(11-12), 1070--1077.

\bibitem{Schlauf1989}
Schlauf, M.; Dimopoulou-Rademann, O.; Rademann, K.; Even, U.; Hensel, F. Direct
  absorption spectra of jet-cooled {Hg$_2$} {\em The Journal of Chemical
  Physics} {\bf 1989}, {\em 90}(8), 4630--4631.

\bibitem{Labanc2018}
Labanc, D.; \v{S}ulka, M.; Pito\v{n}\'{a}k, M.; \v{C}ernu\v{s}\'{a}k, I.;
  Urban, M.; Neogr\'{a}dy, P. Benchmark ccsd(t) and dft study of binding
  energies in be7 - 12: in search of reliable dft functional for beryllium
  clusters {\em Molecular Physics} {\bf 2018}, {\em 116}(10), 1259--1274.

\bibitem{Jiang2012}
Jiang, W.; DeYonker, N.~J.; Wilson, A.~K. Multireference character for 3d
  transition-metal-containing molecules {\em Journal of Chemical Theory and
  Computation} {\bf 2012}, {\em 8}(2), 460--468.

\bibitem{Wang2015}
Wang, J.; Manivasagam, S.; Wilson, A.~K. Multireference character for 4d
  transition metal-containing molecules {\em Journal of Chemical Theory and
  Computation} {\bf 2015}, {\em 11}(12), 5865--5872.

\bibitem{Coe2015}
Coe, J.~P.; Paterson, M.~J. Investigating multireference character and
  correlation in quantum chemistry {\em Journal of Chemical Theory and
  Computation} {\bf 2015}, {\em 11}(9), 4189--4196.

\bibitem{Staroverov2000}
Staroverov, V.~N.; Davidson, E.~R. Distribution of effectively unpaired
  electrons {\em Chemical Physics Letters} {\bf 2000}, {\em 330}(1-2),
  161--168.

\bibitem{NIST_ASD}
Kramida, A.; {Yu.~Ralchenko}.; Reader, J.; {and NIST ASD Team}. {NIST Atomic
  Spectra Database (ver. 5.12), [Online]. Available:
  {\tt{https://physics.nist.gov/asd}} [2025, September 11]. National Institute
  of Standards and Technology, Gaithersburg, MD.}, {\bf 2024}.

\end{thebibliography}
\bibliographystyle{jpc_titles}
\end{document}